\documentclass[aps,prc,twocolumn,showpacs,groupedaddress]{revtex4}

\usepackage{graphicx}

\usepackage{amsmath}

\def\pu1 {$^{241}$Pu$^*$\mbox{ }}

\newcommand{\etal}{{\it et al.}}
\newcommand{\gray}{$\gamma$-ray}
\newcommand{\grays}{$\gamma$ rays}
\newcommand{\g}{$\gamma$}

\newcommand{\CGMF}{$\mathtt{CGMF}$}

\newcommand{\nubar}{$\langle\nu\rangle$}

\begin{document}

\title{Late Time Emission of Prompt Fission Gamma Rays}
\author{P. Talou}
\email{talou@lanl.gov}
\author{T. Kawano}
\author{I. Stetcu}
\author{J. P. Lestone}
\author{E. McKigney}
\author{M. B. Chadwick}
\affiliation{Los Alamos National Laboratory, Los Alamos, NM 87545, USA}

\begin{abstract}
The emission of prompt fission \grays~within a few nanoseconds to a few microseconds following the scission point is studied in the Hauser-Feshbach formalism applied to the deexcitation of primary excited fission fragments. Neutron and \gray~evaporations from fully accelerated fission fragments are calculated in competition at each stage of the decay, and the role of isomers in the fission products, before $\beta$-decay, is analyzed. The time evolution of the average total \gray~energy, average total \gray~multiplicity, and fragment-specific \gray~spectra, is presented in the case of neutron-induced fission reactions of $^{235}$U and $^{239}$Pu, as well as spontaneous fission of $^{252}$Cf. The production of specific isomeric states is calculated and compared to available experimental data. About 7\% of all prompt fission \grays\ are predicted to be emitted between 10 nsec and 5 $\mu$sec following fission, in the case of $^{235}$U and $^{239}$Pu $(n_{\rm th},f)$ reactions, and up to 3\% in the case of $^{252}$Cf spontaneous fission. The cumulative average total \gray~energy increases by 2 to 5\% in the same time interval. Finally, those results are shown to be robust against significant changes in the model input parameters.
\end{abstract}

\date{\today}

\pacs{24.60.Dr, 24.75.+i, 25.85.Ca, 25.85.Ec}

\maketitle

\section{Introduction}

It is common practice to define {\it prompt} fission \grays~as those emitted in {\it coincidence} with a fission event. In actual experiments, the time coincidence window varies from a few nanoseconds to tens, even hundreds of nanoseconds. Theoretical models instead often refer to prompt fission \grays~as those emitted by the fission fragments, before any $\beta$-decay process takes place. This distinction is crucial when comparing experimental data with calculated results. In particular, the presence of isomeric states in the fission fragments complicates this comparison and the proper use of \gray~data in transport simulations.

The proper treatment of the late prompt \gray~emissions is important for several reasons. As discussed by Granier~\cite{Granier:2015}, the estimation of the neutron detector efficiency in a recent experiment~\cite{Chatillon:2014} aimed at measuring precisely the prompt fission neutron spectrum of $^{239}$Pu, was biased due to an improper determination of the time of fission caused by late isomeric decays. In addition, late \grays~can lead to systematic errors in measuring the average prompt neutron multiplicity \nubar~in a large liquid scintillator tank~\cite{Boldeman:1977}. Their study is also of interest for nuclear structure studies~\cite{Urban:2009} in the neutron-rich part of the nuclear chart, as well as for inferring fission fragment yields from estimates of isomeric ratios~\cite{Stetcu:2013}.

In the seventies, isomeric \grays~have been measured in the case of low-energy fission reactions on $^{235}$U and $^{239}$Pu~\cite{Sund:1974} and for $^{252}$Cf(sf)~\cite{John:1970}. Clark \etal~\cite{Clark:1973} also reported the observation of $K$ X rays and \grays~from isomeric decays in the spontaneous fission of $^{252}$Cf. The development of advanced \gray~detection setups such as EUROGAM~\cite{Nolan:1990}, GAMMASPHERE~\cite{Macchiavelli:1998} and EUROBALL~\cite{EUROBALL} led to extensive spectroscopic studies on the fission fragments produced in various thermal neutron-induced fission reactions on actinides~\cite{Gautherin:1998,Gautherin:1997,Urban:2009,Ramayya:1998}. The study of isomers in fission fragments was further pursued to better understand nuclear structure and the onset of nuclear deformation in neutron-rich nuclei. 

The present publication reports on theoretical calculations performed using the Monte Carlo Hauser-Feshbach code \CGMF~\cite{Kawano:2010,Talou:2011} used to describe the statistical decay of excited primary fission fragments. By following the sequential emissions of prompt fission neutrons and \grays, on an event-by-event basis, \CGMF~can be used to study detailed and exclusive characteristics of prompt fission \grays. In particular, the known nuclear structure of fission fragments, including isomeric states, is used to compute the time evolution of \gray~data.

After reviewing some important aspects of the theoretical models and input parameters used in \CGMF~in the next section, we discuss numerical results obtained in the case of neutron-induced fission reactions on $^{235}$U and $^{239}$Pu, and in the spontaneous fission of $^{252}$Cf. 

\section{Theoretical Model}

The Hauser-Feshbach statistical theory of nuclear reactions~\cite{Hauser:1952} is used here to describe the deexcitation of the primary fission fragments by evaporation of neutrons and photons. A Monte Carlo version of this theory has been  implemented in the \CGMF~code~\cite{Kawano:2010,Talou:2011} to follow this deexcitation stage on an event-by-event basis. The Monte Carlo technique is very powerful and straightforward to infer correlations among the many fission observables that are produced in a single fission event. It is also particularly suited to the present study of fission fragment isomeric decays, since specific fragments and \g~lines can be identified and tagged to infer very exclusive data.

In the current work, we assume that all prompt neutrons are emitted from fully accelerated fragments, an assumption with no bearing on the results presented here. At each stage of the decay, the emission probabilities for neutron and photon are sampled in competition with each other. For high nuclear excitation energies, the neutron emission probabilities are much higher than those for \grays. At lower energies, and with the help of high-spin values in the fragments, \gray~emissions become more competitive and probable. 

Low-energy neutron-induced and spontaneous fission reactions of actinides typically produce an asymmetric fission fragment mass yield, with distinct odd-even fluctuations in the charge distribution. In \CGMF, the primary fission fragment yields in mass, charge and kinetic energy, $Y(A,Z,KE)$, are sampled using the Monte Carlo technique. An underlying assumption of this work is that we know the pre-neutron fission fragment yields accurately enough. This assumption is certainly valid in the case of the spontaneous fission of $^{252}$Cf, and to some extent to the thermal neutron-induced fission yields of $^{235}$U and $^{239}$Pu. While no experimental data exist on the complete yields in mass, charge and kinetic energy, we had to reconstruct those yields based on a combination of experimental data and systematics. At higher incident neutron energies, the situation is much more complex and uncertain, especially as we move across the multiple-chance fission thresholds where one or more neutrons are emitted prior to the fission of the residual compound nucleus. Most experimental data on the mass yields come from 2$E$ experiments~\cite{Budtz:1988}, in which the kinetic energies of both fragments are measured, with a typical mass resolution of 4-5 amu. It can also be significantly impacted by our lack of knowledge of the fragment mass-dependent average neutron multiplicity as a function of the incident neutron energy, $\langle \nu \rangle(A,E_{inc})$. Since the population of specific isomers depends on the initial yields, the present results will be somewhat sensitive to uncertainties in this initial input. 

For a specific fragmentation into a light and a heavy fragments, the energy release $E_r$ is calculated using experimentally known nuclear masses from the Audi-Wapstra 2012 table~\cite{Audi:2012} and using those calculated by M\"oller \etal~\cite{Moller:2015} using the Finite-Range Droplet Model (FRDM) for nuclei whose mass has not been measured. The total excitation energy $TXE$ is then inferred as $E_r-TKE$, where $TKE$ is the total kinetic energy of the fragments obtained from experimental data. The total excitation energy $TXE$ includes both intrinsic and collective energy components, and is eventually evaporated through neutron and \gray~emissions. Knowing $TXE$ is not enough to start the decay of the fragments as it needs to be shared between the two fragments. Our default calculations share this energy in order to best reproduce the well-known ``saw-tooth" curve for the mass-dependent average neutron multiplicity, \nubar$(A)$. However, we do study the sensitivity of our results to this particular prescription, as discussed in Section~\ref{sec:results}.

Various studies~\cite{Huizenga:1960,Wilhelmy:1972,Naik:2005,Stetcu:2014} have shown that the initial spin distribution in the fragments typically averages over 6-8$\hbar$. Such rather high, model-dependent values are needed to explain isomeric ratio measurements as well as prompt \gray~data that indicate a competition between neutrons and \grays~at excitation energies significantly higher than the average neutron separation energy. In \CGMF, the initial spin distribution is given by
\begin{eqnarray}
P(J) \propto (2J+1)\exp \left[ -\frac{J(J+1)}{2B^2(A,Z,T)}\right],
\end{eqnarray}
where the spin cut-off parameter $B^2$ is
\begin{eqnarray}
B^2(A,Z,T) = \alpha \frac{\mathcal{I}_0(A,Z)T}{\hbar^2},
\end{eqnarray}
with $T$ the fragment temperature, and $\mathcal{I}_0$ the ground-state rigid-body moment of inertia of the fragment $(A,Z)$
\begin{eqnarray}
\mathcal{I}_0 &=& \frac{2}{5} A^{5/3}r_0^2\frac{(M_n+M_p)c^2}{2} \\ \nonumber
&\times& \left( 1.0+0.32\beta_2(Z,A)+0.44\beta_2^2(Z,A)\right).
\end{eqnarray}
In this expression, the first line represents the moment of inertia of a solid sphere, while the second line represents the impact of an ellipsoidal deformation through the $\beta_2$ quadrupole deformation parameter. In the present work, $\beta_2$ values are taken from the FRDM 2012 calculations~\cite{Moller:2015}.

$\alpha$ is a global scaling parameter that is adjusted to best reproduce the observed average prompt fission \gray~spectrum~\cite{Stetcu:2014}. Typically, $\alpha$ is set to 1.5$-$1.7. The sensitivity of our results to this parameter is presented in Section~\ref{sec:results}.

The Hauser-Feshbach statistical theory of nuclear reactions~\cite{Hauser:1952} states that the cross section for the reaction $X(a,b)Y$ is given by
\begin{eqnarray} \label{eq:HF}
\sigma_{ab} = \sigma_{a,CN} \times \frac{T_b}{\sum_c T_c},
\end{eqnarray}
where $T_c$ represents the transmission coefficient for the decay in the reaction channel $c$. In the description of the deexcitation of the fission fragments, the incident channel $a$ is only implicitly considered, since the fission fragments are already formed, and only the deexcitation part is treated explicitly. In low-energy fission reactions, the fragments are formed with moderate excitation energies, typically less than 30 MeV, and only two decay channels dominate: neutron and gamma. The emission of other light-charged particles is hindered by the Coulomb barrier. Note that the special case of ternary fission is not accounted for in this study.

Applying Eq.~(\ref{eq:HF}) to the decay of excited fission fragments, we are left with calculating the transmission coefficients for neutron emission, $T_n$, and $\gamma$ decay, $T_\gamma$. Considering not only discrete final states but also a continuum description of final states, the probability to emit a neutron of energy $\epsilon_n$ is proportional to the product of the energy-dependent transmission coefficients $T_n(\epsilon_n)$ and the level density in the residual nucleus at the final excitation energy $(E-\epsilon_n-S_n)$:
\begin{eqnarray}
P_n(\epsilon_n)d\epsilon_n \propto T_n(\epsilon_n)\rho(Z,A-1,E-\epsilon_n-S_n)d\epsilon_n.
\end{eqnarray}
In this equation, $S_n$ is the neutron separation energy of the residual nucleus. The neutron transmission coefficients are obtained as usual from the scattering matrix $S$ as $T_c=1-|\langle S_{cc}\rangle|^2$. The global spherical optical model of Koning and Delaroche~\cite{Koning:2003} is used in the present work. Many of the fragments produced in a fission reaction are deformed, and the use of a spherical optical model is certainly not perfectly adapted to the present calculations. In particular, the average prompt fission neutron spectrum has been traditionally calculated too soft by \CGMF. The small under-prediction of the neutron spectrum tail (above $\sim$ 5 MeV) should have only a marginal effect on the \g~results discussed in this work though.

The neutron transmission coefficients also depend on the energy and spin of the emitted neutron, and are more explicitly denoted by $T_n^{lj}(\epsilon_n)$. Energy, spin and parity conservation rules apply at every step of the evaporation process, and 
\begin{eqnarray}
E_f&=&E_i-\epsilon, \nonumber \\
|J_i-j| \leq &J_f& \leq J_i + j, \nonumber \\
\pi_f &=& \pi_i(-)^l,
\end{eqnarray}
where the $i$ and $f$ indices indicate initial and final states, respectively. The energy $\epsilon$ represents the energy of the ejectile, i.e., $\epsilon_n+S_n$ in the case of a neutron, with $S_n$ the neutron separation energy, and simply $\epsilon_\gamma$ in the case of a \g~ray.

Similarly, the probability to emit a photon with energy $\epsilon_\gamma$ is given by
\begin{eqnarray}
P_\gamma(\epsilon_\gamma)d\epsilon_\gamma \propto T_\gamma(\epsilon_\gamma)\rho(Z,A,E-\epsilon_\gamma)d\epsilon_\gamma.
\end{eqnarray}
The \gray~transmission coefficients are obtained as usual from the strength functions $f_{Xl}(\epsilon_\gamma)$ inferred from giant-dipole resonances as
\begin{eqnarray}
T_\gamma(\epsilon_\gamma) = 2\pi\epsilon_\gamma^{2l+1}f_{Xl}(\epsilon_\gamma),
\end{eqnarray}
where $Xl$ represents the multipolarity of the transition. Only $E1$, $M1$ and $E2$ transitions are considered in this work, with parameterizations taken from the RIPL-3 database~\cite{RIPL3}. For $E1$ transitions, the generalized Lorentzian form of Kopecky and Uhl~\cite{Kopecky:1990} is used. A standard Lorentzian form is used for both $M1$ and $E2$ transitions. In this picture, $E1$ transitions strongly dominate in the continuum, while collective $E2$ transitions strongly reduce the fragment angular momentum following the $Y_{\rm rast}$ line.

Another important ingredient in the equations considered above is the nuclear level density $\rho(Z,A,U,J,\pi)$ in excitation energy $U$, spin $J$ and parity $\pi$. The phenomenological model of Gilbert and Cameron~\cite{Gilbert:1965} is used for all fission fragments. In this model, the low-energy part of the density is represented by a constant-temperature regime and connected at the lowest energies to known discrete states. At higher energies, a Fermi-gas representation is used instead and smoothly connected to the constant-temperature formula. As usual, the parameters are adjusted to match the cumulative number of known discrete states at the lowest energies, and to match the average level spacing measured at the neutron separation energy. For neutron-rich fission fragments however, those data are not as well known as for nuclei near the valley of stability, and one has to rely on systematics, which necessarily increase the uncertainties associated with the calculated results.

Known discrete states are taken from the RIPL-3 library~\cite{RIPL3}, which very closely matches the ENSDF nuclear structure data library~\cite{ENSDF}. In the discrete region, known branching ratios are also used to follow the decay of particular states. Of particular interest for the present study is the existence of isomeric states in the fission fragments, which may delay the emission of prompt \grays. The \CGMF~code uses the known half-lives of those isomers to determine if the decay should take place or not, within the chosen time coincidence window. It is important to realize that \CGMF~calculations do not predict the low-lying structure of the fission fragments and therefore do not predict the presence of or lack of an isomer. However, \CGMF\ does calculate the population of these isomers based on the initial fission fragments and the calculated neutron probability distributions.

\section{Numerical Results} \label{sec:results}

\subsection{General Results on Prompt Neutrons and Gamma Rays}

Before addressing the central question of the time-dependence of the prompt \grays~that interests us in this paper, it is important to study the accuracy of the general results obtained using \CGMF~on all prompt fission \grays. The spontaneous fission of $^{252}$Cf and the thermal neutron-induced fissions of $^{235}$U and $^{239}$Pu provide very good validation points as quality experimental data exist for those isotopes and reactions.

\begin{figure}[ht]
\centerline{\includegraphics[width=\columnwidth]{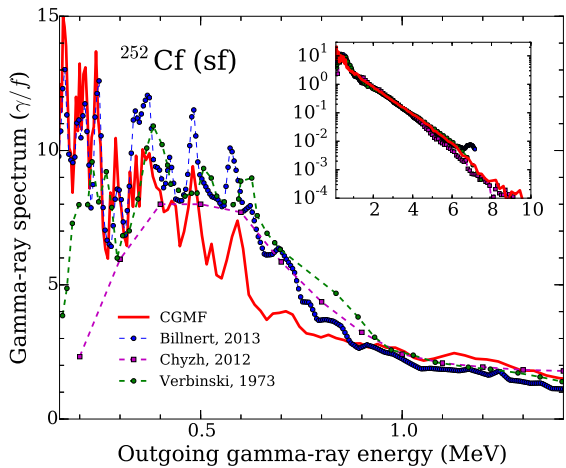}}
\caption{\label{fig:Cf252sf-pfgs}Average prompt fission \gray~spectrum (PFGS) for the spontaneous fission of $^{252}$Cf. The high-energy tail of the calculated PFGS reproduces very well the recent measurement by Billnert \etal~\cite{Billnert:2013} (see inset), while the low-energy part of the spectrum shows very clear structures that can be attributed to transitions between low-lying excited states in fission fragments. \CGMF~calculations do reproduce many of these structures reasonably well. The exact magnitude of these peaks depends on several factors, including our present knowledge of the nuclear structure of fission fragments. Note that no energy resolution broadening has been applied to the \CGMF~results.}
\end{figure}

\begin{figure}[ht]
\centerline{\includegraphics[width=\columnwidth]{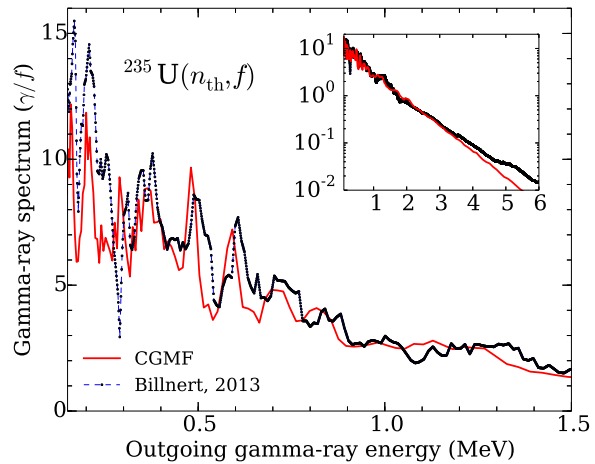}}
\caption{\label{fig:U235T-pfgs}Same as Fig.~\ref{fig:Cf252sf-pfgs} for the thermal neutron-induced fission reaction of $^{235}$U. Experimental data are from Billnert \etal~\cite{Billnert:2013}. The observed low-energy structures, as shown in the inset, are very well reproduced by \CGMF~calculations. The high-energy tail would be better fit with a smaller value for the $\alpha$ spin parameter.}
\end{figure}

\begin{figure}[ht]
\centerline{\includegraphics[width=\columnwidth]{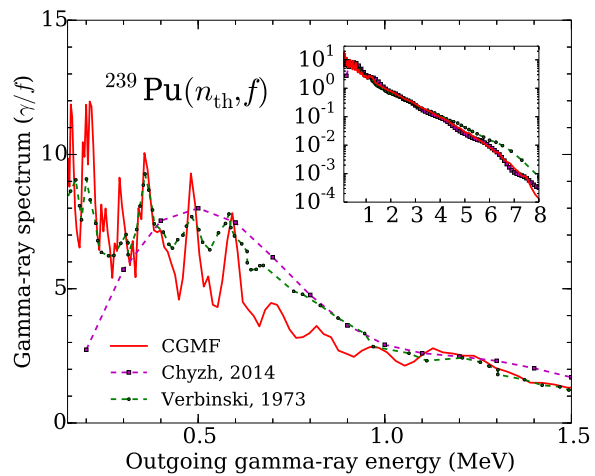}}
\caption{\label{fig:Pu239T-pfgs}Same as Fig.~\ref{fig:Cf252sf-pfgs} for the thermal neutron-induced fission reaction of $^{239}$Pu.}
\end{figure}

Figures~\ref{fig:Cf252sf-pfgs}--\ref{fig:Pu239T-pfgs} show the average prompt fission \gray~spectrum (PFGS) for those three reactions respectively. The overall agreement between the calculated and experimental PFGS is remarkable. In particular, the very strong fluctuations of the \g~spectrum observed below 1 MeV are nicely reproduced by the calculations for the most part. Older experimental data by Verbinski \etal~\cite{Verbinski:1973} do exhibit similar fluctuations, but with a rather low energy resolution. Recent higher-energy resolution experiments by Billnert \etal~\cite{Billnert:2013} performed with LaBr$_3$ and CeBr$_3$ detectors reported much more detailed structures, consistent with Verbinski data. The recent results obtained by Chyzh \etal~\cite{Chyzh:2013,Chyzh:2014} used the DANCE calorimeter at LANL to infer PFGS and multiplicity-dependent spectra. A Bayesian inference scheme was used to recover the PFGS from the raw measured data. Not surprisingly, no detailed structure could be found following this approach, but the overall tail of the PFGS should be relatively trustworthy. Note that the \CGMF~results include a Doppler correction for the \gray~energies, but do not include any energy resolution that would further broaden the calculated peaks.

In general, our calculated results are compatible with all these experimental results. The structures observed in the thermal neutron-induced fission of $^{235}$U and $^{239}$Pu (see Figs.~\ref{fig:U235T-pfgs} and~\ref{fig:Pu239T-pfgs}) are particularly well reproduced by \CGMF~calculations. Some discrepancies remain in the exact magnitude of those peaks. The calculated magnitudes depend on three key factors: the primary fission fragment yields, the neutron emission probability distribution for each initial fission fragment, and the low-lying nuclear structure information (energy levels, branching ratios, half-lives) contained in the nuclear structure database~\cite{RIPL3} used in the \CGMF~calculations. A thorough spectroscopic investigation is underway to determine which of these three factors can explain some of the observed discrepancies. Obviously, erroneous isomeric half-lives would impact the results presented below.

\begin{figure}[ht]
\centerline{\includegraphics[width=0.75\columnwidth]{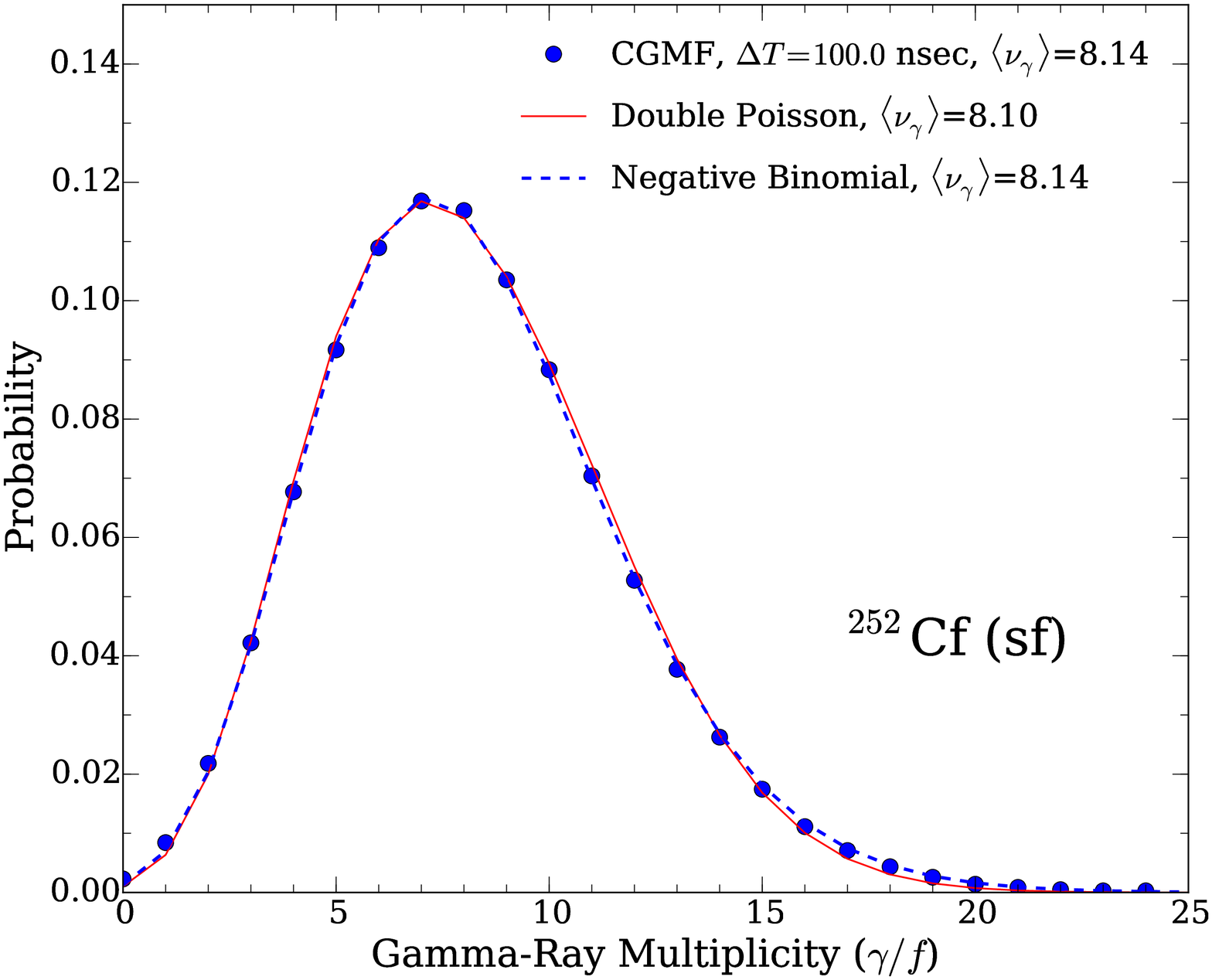}}
\centerline{\includegraphics[width=0.75\columnwidth]{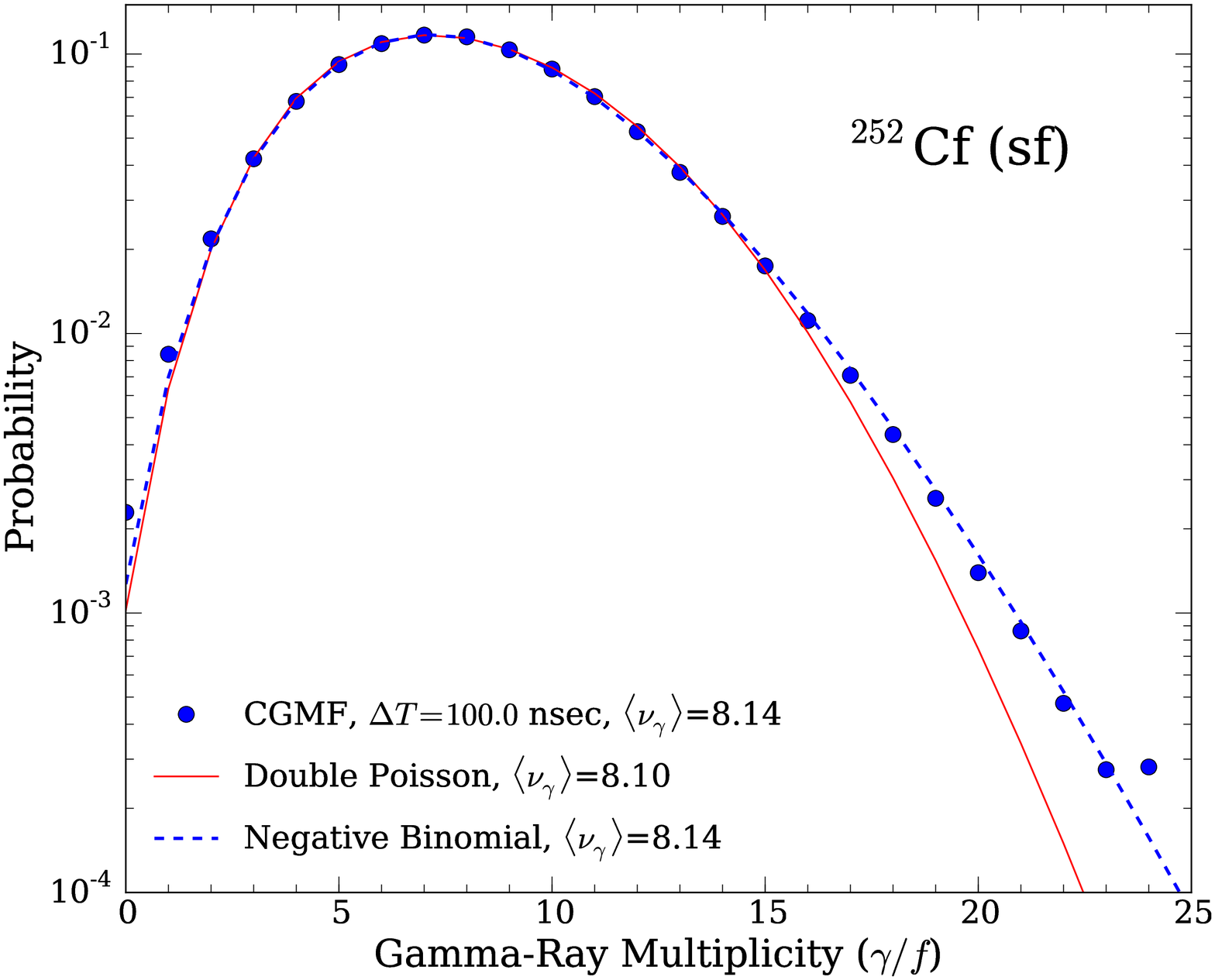}}
\caption{\label{fig:Cf252sf-Pnug}(top) Prompt fission \gray~multiplicity distribution calculated with \CGMF~and to compared to a negative binomial distribution~\cite{Valentine:2001} with best-fit parameters. A double Poisson distribution as proposed originally by Brunson~\cite{Brunson:1982} to fit his experimental results is shown in red on the log-plot (bottom). The negative binomial distribution is in much better agreement with the tail of the distribution as calculated by \CGMF~than the double Poisson distribution. The \gray~threshold used in the \CGMF~results is 140 keV.}
\end{figure}

The calculated prompt \gray~multiplicity distribution $P(\nu_\gamma)$ is shown in Fig.~\ref{fig:Cf252sf-Pnug}. Brunson~\cite{Brunson:1982} used a double Poisson distribution with three parameters to fit his experimental results. Those parameters did not contain any explicit physical information, but were found to best fit the experimental data only. Valentine~\cite{Valentine:2001} used instead a negative binomial distribution with only two parameters. Both types of distribution can be used to represent an over-dispersed Poisson distribution, where the width is larger than the mean, as in the case of the prompt fission \gray~probability distribution. Using the negative binomial distribution, the probability for emitting $n$ \grays~is
\begin{eqnarray}
P(n)=\binom{\alpha+n-1}{n}p^\alpha (1-p)^n,
\end{eqnarray}
with the distribution parameters $p=\langle n\rangle/\sigma^2$ and $\alpha=p\overline{n}/(1-p)$. Those parameters can be expressed as
\begin{eqnarray}
\alpha=\left(D_\gamma-1\right)^{-1} \mbox{ ; } p=\frac{\alpha}{\alpha+\langle \nu_\gamma\rangle},
\end{eqnarray}
where $D_\gamma$ is the relative width defined as
\begin{eqnarray}
D_\gamma = \frac{\langle n(n-1)\rangle}{\langle n\rangle^2}.
\end{eqnarray}
A log-plot (see bottom plot of Fig.~\ref{fig:Cf252sf-Pnug}) of the $P(\nu_\gamma)$ distribution calculated with \CGMF~shows that the predicted tail is much better represented by the negative binomial distribution, with $D_\gamma = 1.06$, than by the best-fit double Poisson distribution. The \CGMF~calculations show an excess of fission events with no \gray~emission, i.e., for $P(\nu_\gamma=0)$, compared to the negative binomial fit. Overall, the agreement between the \CGMF~results and the negative binomial distribution is remarkable, up to $\nu_\gamma\simeq 24$, over almost 3 orders of magnitude. Similar results were obtained in the case of $^{239}$Pu and $^{235}$U $(n_{\rm th},f)$ reactions. Note that the multiplicity distribution as well as other \gray~observables are very sensitive to the threshold used for the photon detection~\cite{Stetcu:2014}.

\subsection{Late Time Emissions of Prompt Gamma Rays}

The general good agreement between experiments and calculations presented above provides strong support for the validity of the model and code used to describe the emission of prompt neutrons and \grays. As mentioned earlier, the known discrete level structure of each fragment is used at the lowest excitation energies, matched to a level density description of a continuum of levels at higher energies. In addition to the energy, spin and parity known (or assumed) for the lowest energy levels, some information exists on the half-lives for some of these levels. If not present, we assume that the decay happens instantaneously; typical compound nucleus half-lives are of the order of 10$^{-19}$ to $10^{-15}$ sec. However, if a half-life is known for a particular excited state, \CGMF~will randomly sample the exponential decay law according to an input time window in coincidence with the fission event, and determines if the decay should occur. In this way, the calculated \g~spectra and multiplicities will vary with the time coincidence window.

\begin{figure}[ht]
\includegraphics[width=\columnwidth]{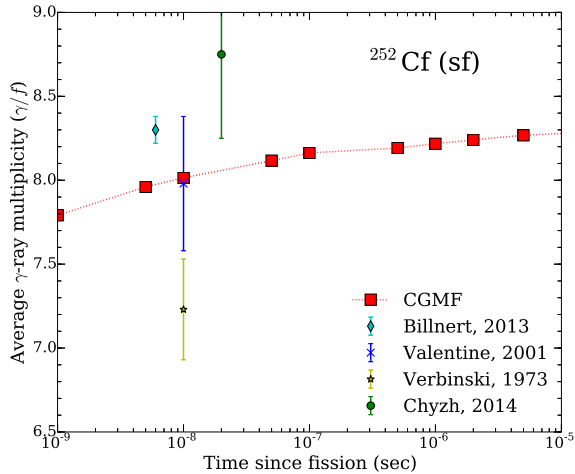}
\caption{\label{fig:Ngt-Cf252SF}Average prompt \gray~multiplicity as a function of time for $^{252}$Cf (sf). Experimental data are from Verbinski~\cite{Verbinski:1973}, Billnert~\cite{Billnert:2013}, and Chyzh~\cite{Chyzh:2014}. Valentine data point~\cite{Valentine:2001} represents a weighted average of experimental data available prior to 2001.}
\end{figure}

In Fig.~\ref{fig:Ngt-Cf252SF}, the average prompt fission \gray~multiplicity, $\langle \nu_\gamma\rangle$, is plotted as a function of time since fission for $^{252}$Cf (sf). This quantity increases by more than 5\% between 1 nsec and 5 $\mu$sec, with $\langle \nu_\gamma\rangle$=7.79 and 8.27, respectively. This evolution is not necessarily linear, as illustrated here, as it depends on the distribution of populated isomer half-lives. The calculated results are compared to average \gray~multiplicity reported in various publications. If one believes the reported experimental error bars, these results are mostly inconsistent. An ``average" value was obtained by Valentine~\cite{Valentine:2001}, with a ``conservative" standard deviation of 2.5\%. Our calculated result is in very good agreement with this value, considering a time coincidence window of 10 nsec.

\begin{figure}[ht]
\centerline{\includegraphics[width=0.75\columnwidth]{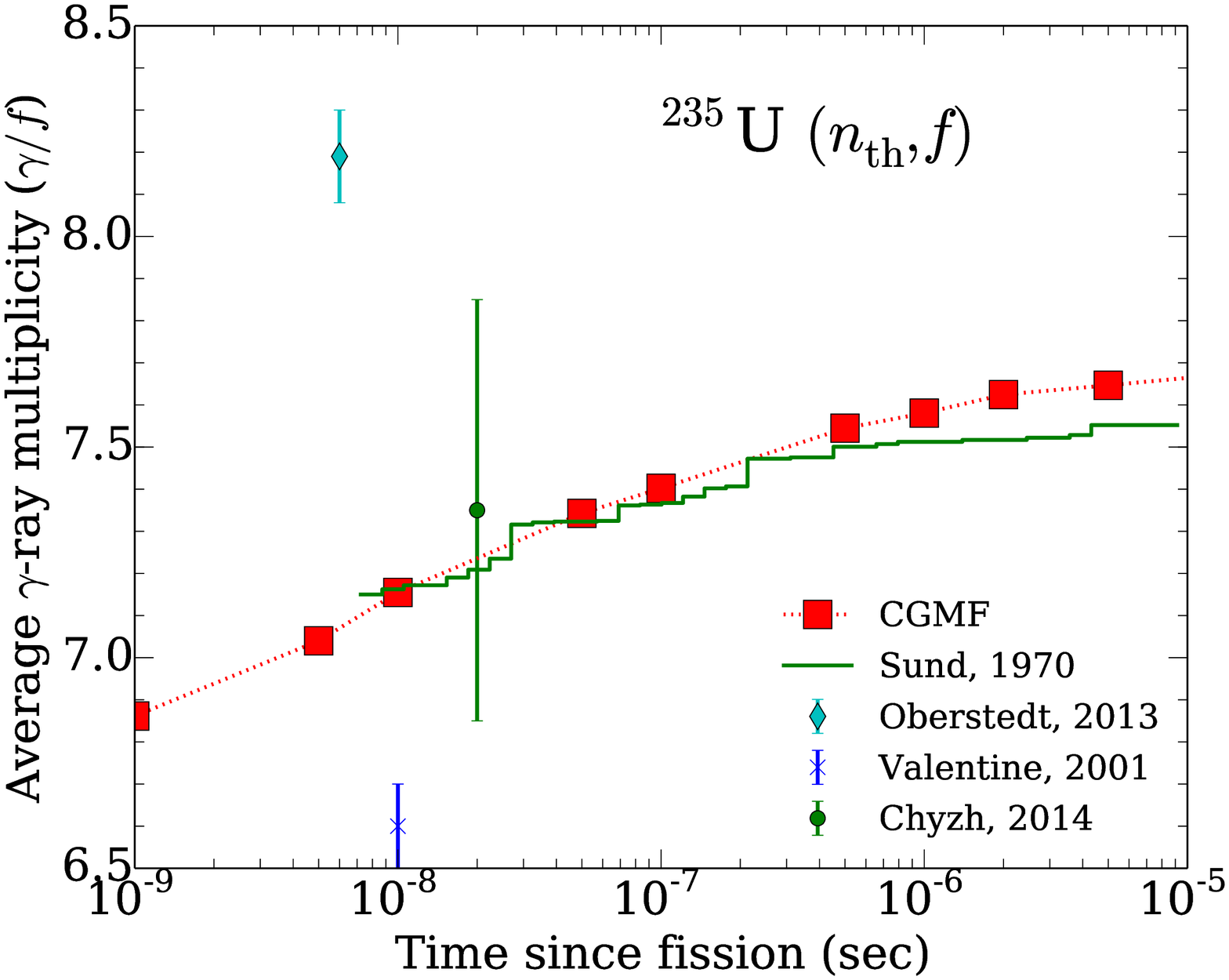}}
\centerline{\includegraphics[width=0.75\columnwidth]{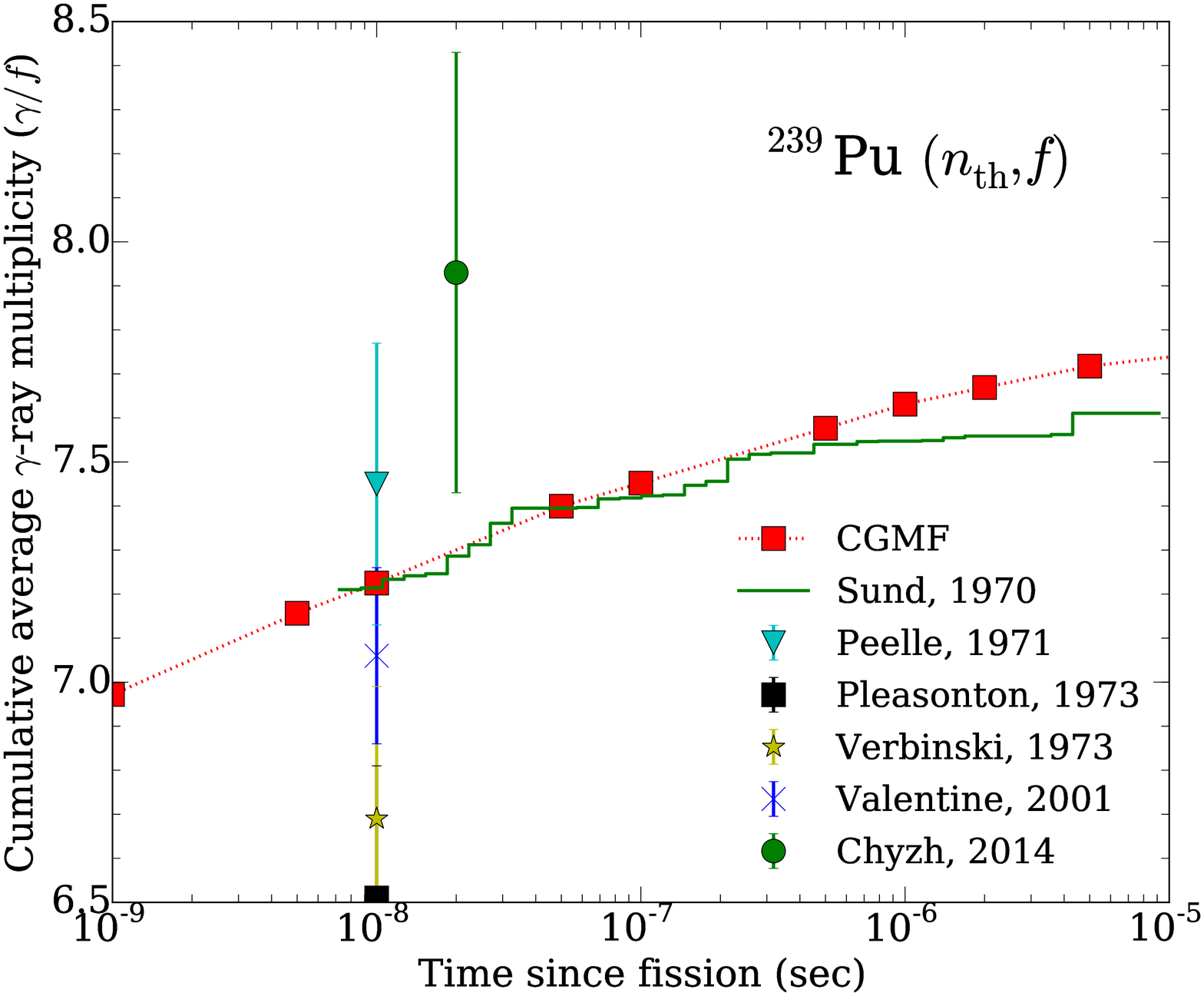}}
\caption{\label{fig:Ngt-U5T-Pu9T}Same as Fig.~\ref{fig:Ngt-Cf252SF} for the thermal neutron-induced fission of $^{235}$U (top) and $^{239}$Pu (bottom). The experimental data by Sund \etal~\cite{Sund:1974} (green solid steps) were normalized to match \CGMF-calculated average \gray~multiplicity at 10 nsec.}
\end{figure}

Similar results were obtained in the thermal neutron-induced fission reactions of $^{235}$U and $^{239}$Pu (Fig.~\ref{fig:Ngt-U5T-Pu9T}), but in this case, the changes are even more important: about 10\% of the prompt \grays~are emitted after 1 nsec. Given the range and uncertainties associated with the different experimental data points, our calculated results are in reasonable agreement with the values reported in the ENDF/B-VII.1 library~\cite{ENDFBVII1}: assuming a time since fission of 10 nsec, \CGMF-calculated average total \gray~energies are 6.32 and 6.55 MeV for $^{235}$U and $^{239}$Pu, respectively, compared to 6.60 and 6.74 MeV in ENDF/B-VII.1. The calculated result for $^{252}$Cf (sf) is 6.73 MeV, in very good agreement with the most recent experimental data by Oberstedt \etal~at 6.65$\pm$0.10 MeV~\cite{Oberstedt:2015}. At this point, it is important to note that evaluated nuclear data libraries do not contain any specific information on the time of emission of the prompt \grays, albeit making a clear distinction between prompt and $\beta$-delayed emissions. A reasonable assumption would be to consider the average prompt \gray~multiplicity as the maximum average number of \grays~emitted before $\beta$-decay takes place, which can be neglected before 1 msec after fission.

Figure~\ref{fig:Ngt-U5T-Pu9T} also shows a comparison with the experimental data obtained by Sund \etal~\cite{Sund:1974} on \grays~emitted between 20 nsec and about 1 $\mu$sec. The half-life and intensity of the delayed \gray~resolved peaks reported in Table I of Ref.~\cite{Sund:1974} were used to compute the cumulative average \gray~multiplicity as a function of time. Although Sund \etal~measured the \grays\ between 20 nsec and about 1 $\mu$sec, they inferred half-lives ranging from a few nsec to a few $\mu$sec. Sund \gray~multiplicities for both isotopes were normalized to \CGMF\ results at 10 nsec. The agreement between \CGMF\ and the experimental data is very good, and the slight overestimation of \CGMF-calculated multiplicity compared to Sund data could easily be accounted for because of missing isomers not identified in this experiment. For times greater than a few $\mu$sec, we expect the discrepancies to increase as Sund experiment was limited to this upper time limit.

\subsection{Fission Fragment Isomers}

\begin{figure}[ht]
\centerline{
\includegraphics[width=\columnwidth]{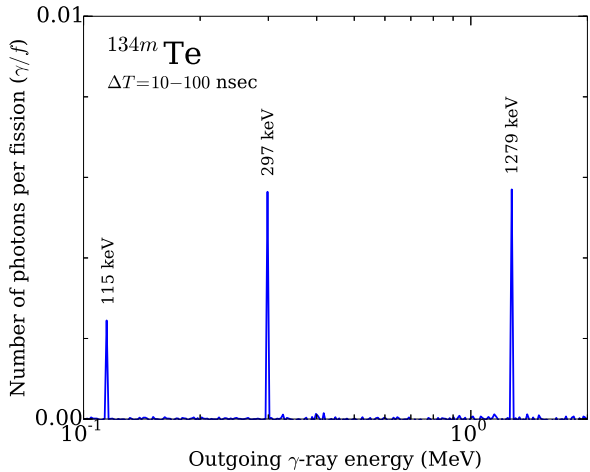}
}
\caption{\label{fig:gSpec-Te134m}Gamma-ray spectrum in the 10-100 nsec coincidence window gated on the post-neutron emission $^{134}$Te fission fragment. The three \g~lines at 115, 297 and 1279 keV corresponding to the deexcitation of the 162 nsec isomer in $^{134}$Te are clearly visible in this time range. The intensity of the 115 keV is strongly hindered compared to the two other lines due to internal conversion.}
\end{figure}

The evolution in time of the cumulative average \gray~energy and multiplicity is governed by the half-lives of the isomers present in fission fragments and by the probability of feeding these isomers in the fission reaction. This latter probability depends in turn on the primary fission fragment yields and on the deexcitation paths for each fragment. A well-known example of an isomeric state in fission fragments is the 162 nsec, 1.69 MeV, $6^+$ isomer in $^{134}$Te, whose decay to the ground-state occurs via 3 \g~lines at 115, 297 and 1279 keV. Figure~\ref{fig:gSpec-Te134m} shows \CGMF~numerical results for the spectrum of \grays~emitted in the 10 to 100 nsec time range following fission, with a gate on mass 134, after neutron emission. By studying the \g~spectrum after 10 nsec, and by focusing on only one fragment mass, the background is almost completely eliminated, and the discrete decay lines appear very strongly. The reduced intensity of the 115 keV line compared to the other two lines is due to internal conversion, which is taken into account in our simulations. This result represents the deexcitation of the isomer after being populated not only through $^{134}$Te, but also through $^{135,136}$Te, with one and two neutron(s) being emitted prior to populating the isomeric state.

\begin{figure}[ht]
\centerline{\includegraphics[width=\columnwidth]{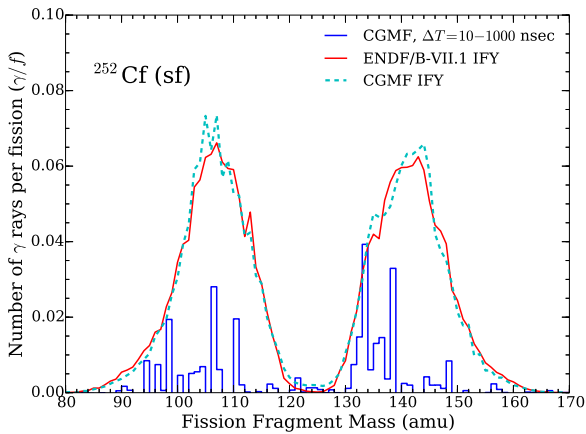}}
\caption{\label{fig:lateGammasPerMass}The distribution of prompt fission \grays~emitted between 10 nsec and 1 $\mu$sec after fission, for $^{252}$Cf spontaneous fission, is shown as the blue histogram. Also shown are the independent fission yields (IFY), i.e., post-neutron emission fission yields, as calculated by \CGMF~(dashed cyan) and taken from the ENDF/B-VII.1 library (solid red), in arbitrary units.}
\end{figure}

The number of prompt \grays~emitted 10 nsec to 1 $\mu$sec following fission is shown in Fig.~\ref{fig:lateGammasPerMass} as a function of the fission fragment mass. For comparison, the independent fission yields (IFY), i.e., fission fragment mass yields after prompt neutron emission, are shown for the ENDF/B-VII.1 evaluated library (solid red) and for the ones resulting from the \CGMF~calculations (cyan dashed). The prompt isomeric decays are concentrated in the 95-110 amu and 130-140 amu mass ranges. We can also note the reasonable agreement between the \CGMF~calculated IFY and the ones evaluated in the ENDF/B-VII.1 library.

The detailed study of specific \g~decay chains is fraught with uncertainties. In particular, the low-lying nuclear structure of each fission fragment has to be known with great accuracy. \CGMF~does not predict level schemes, but instead relies on what is reported in modern nuclear structure databases, which reflects our current state of knowledge. While many spectroscopic studies with modern \gray~detector arrays have improved this knowledge significantly, it is also clear that quite of few uncertainties remain. If a particular isomeric state is missing in the database, \CGMF\ would not predict it either. As an example, the partial level scheme obtained~\cite{Nowak:1999} in $^{139}$Cs indicates the presence of a (15/2)$^+$ level at 1145.8 keV, decaying through the (11/2)$^+$ state at 601.5 keV, and finally to the 7/2$^+$ ground-state. Instead, the evaluated nuclear structure database ENSDF~\cite{ENSDF} only lists the energy of this state, without further information on its spin, parity and decay scheme. In the current RIPL-3 database~\cite{RIPL3}, this level is shown as (11/2)$^-$ with a direct decay to the ground-state. In yet another preliminary version of the RIPL-3 level scheme, this state is shown as an (11/2)$^+$ state with a direct feeding to the ground-state. It is not our aim here to decide which one is correct, but instead to point out that such uncertainties can have a large impact on the present numerical predictions for specific fragments. Average quantities however are much less sensitive to the details of the nuclear structure of each fragment.

\subsection{Late Gamma-Ray Energy Spectrum and Multiplicity}

\begin{figure}[ht]
\centerline{\includegraphics[width=\columnwidth]{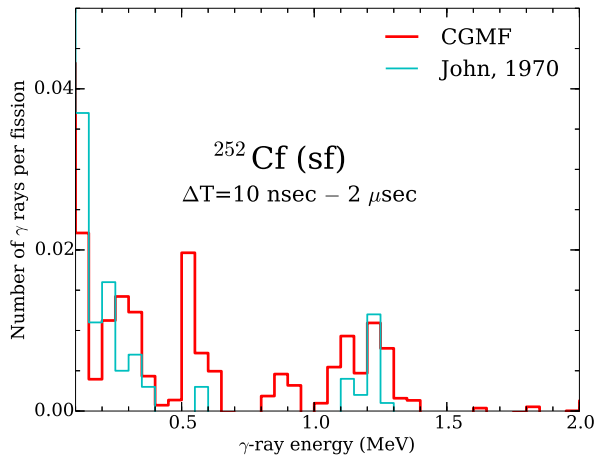}}
\caption{\label{fig:Cf252-pfgs-10-2000}Energy spectrum of late fission \grays~emitted in the 10 nsec to 2 $\mu$sec time window following fission, in the case of $^{252}$Cf spontaneous fission. Experimental data are from John~\etal~\cite{John:1970}.}
\end{figure}

It is also instructive to analyze the energy spectrum of the late prompt \grays. \CGMF~results are shown in Fig.~\ref{fig:Cf252-pfgs-10-2000} in the case of \grays~emitted in the 10 nsec to 2 $\mu$sec time range following the spontaneous fission of $^{252}$Cf, and compared with experimental data by John \etal~\cite{John:1970}. The figure shows the spectrum up to 2.0 MeV only, as almost no intensity is observed at higher energies, as already noted by Sund~\cite{Sund:1974} who measured up to 4 MeV. The calculated results are in relatively good agreement with the data of John \etal~\cite{John:1970}. \CGMF\ predicts relatively more late \grays\ than reported by John, a result somewhat expected since many more isomers where discovered after John and Sund experiments with the advent of high-resolution \gray~detectors such as GAMMASPHERE or EUROGAM. Thanks to these detectors, many $\gamma-\gamma$ and $\gamma-\gamma-\gamma$ coincidences were used to remove most of the background, and uncover the existence of additional isomers in many fission fragments. Also note that \CGMF~reproduces the peak observed near 1.3 MeV, corresponding to the prominent production of isomers in Te and Xe isotopes.

\begin{figure}[ht]
\centerline{\includegraphics[width=\columnwidth]{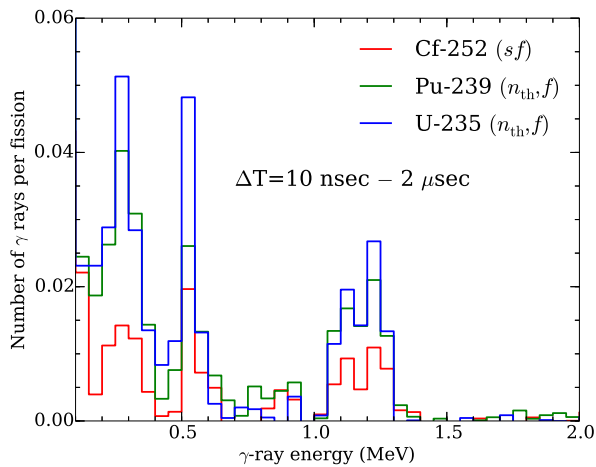}}
\caption{\label{fig:pfgs-10-2000}Same as Fig.~\ref{fig:Cf252-pfgs-10-2000} but for all three fission reactions studied.}
\end{figure}

We performed similar calculations for the thermal neutron-induced fission of $^{239}$Pu and $^{235}$U, as shown in Fig.~\ref{fig:pfgs-10-2000}, in the same time window of 10 nsec to 2 $\mu$sec. While the overall shapes of the calculated spectra are similar, noticeable differences appear in places. The late \g~spectrum for $^{235}$U is significantly more pronounced near the 0.5 and 1.3 MeV peaks, as well as near 0.2 MeV. 

\begin{figure}[ht]
\centerline{\includegraphics[width=\columnwidth]{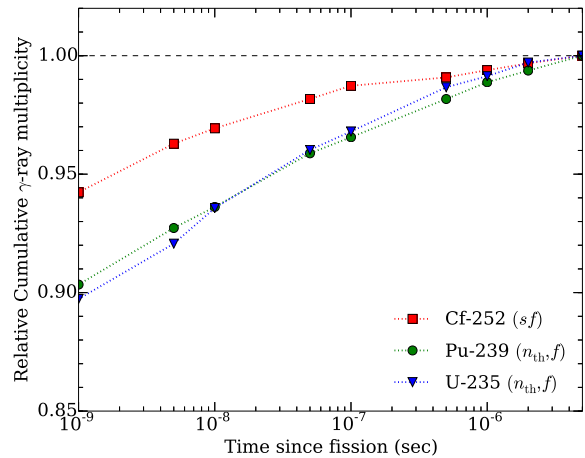}}
\caption{\label{fig:Ngt-all}Cumulative average prompt \gray~multiplicity as a function of time since fission for thermal neutron-induced fissions of $^{235}$U and $^{239}$Pu, and spontaneous fission of $^{252}$Cf. }
\end{figure}

\begin{figure}[ht]
\centerline{\includegraphics[width=\columnwidth]{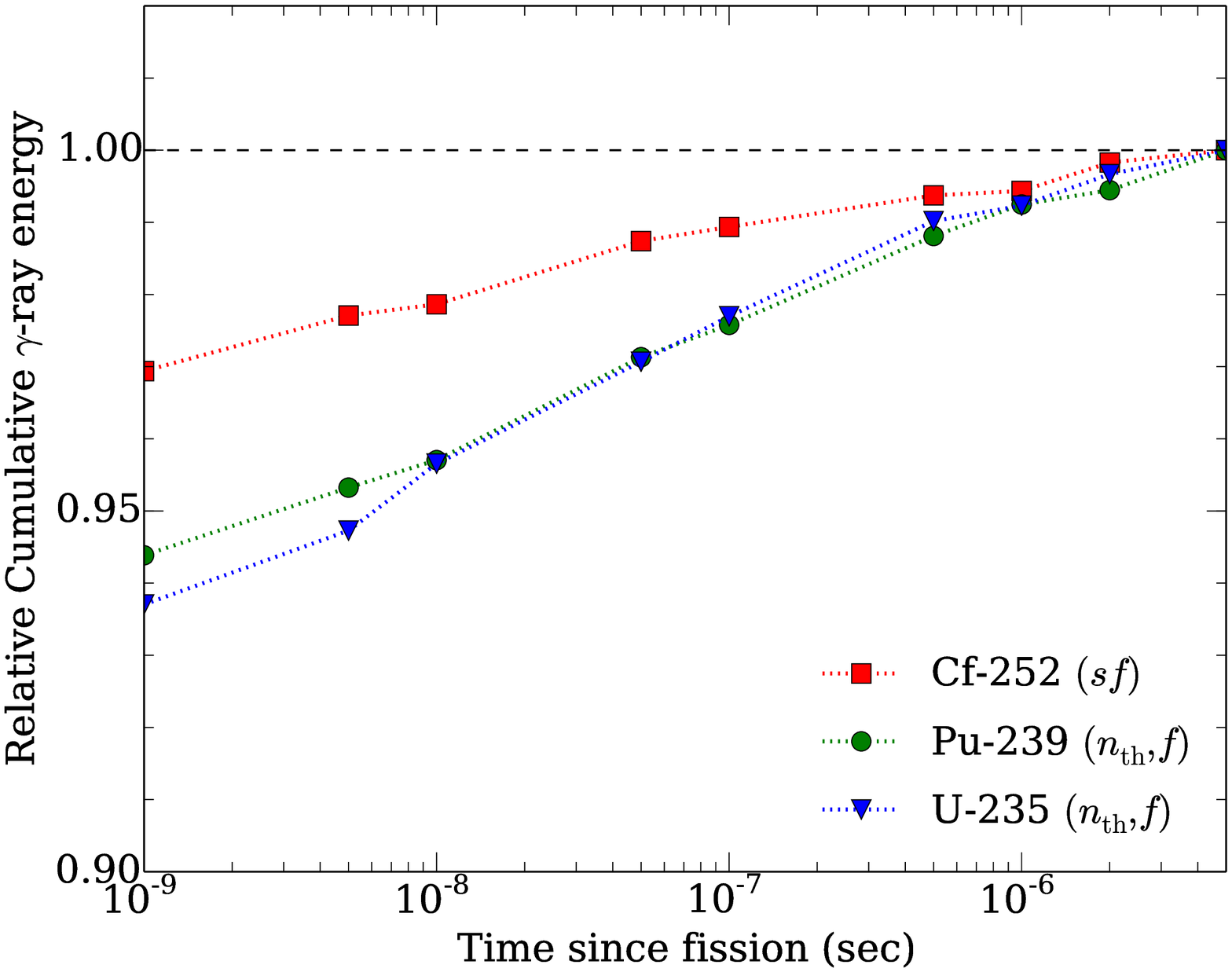}}
\caption{\label{fig:Egt-all}Cumulative average prompt total \gray~energy as a function of time since fission for thermal neutron-induced fissions of $^{235}$U and $^{239}$Pu, and spontaneous fission of $^{252}$Cf.}
\end{figure}

\begin{figure}[ht]
\centerline{\includegraphics[width=\columnwidth]{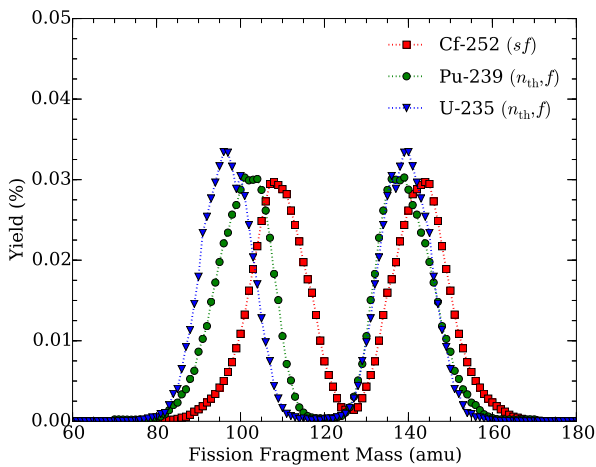}}
\caption{\label{fig:YA}Pre-neutron emission fission fragment mass yields for the three reactions studied here.}
\end{figure}

Distinctions between the time evolutions of the late \gray~emissions in the three fission reactions studied are summarized in Figs.~\ref{fig:Ngt-all} and~\ref{fig:Egt-all} for the average cumulative \gray~multiplicity and total \gray~energy, respectively. The slopes of $\langle N_\gamma\rangle(t)$ are steeper and very similar for $^{235}$U and $^{239}$Pu than for $^{252}$Cf. A similar observation can be made for $\langle E_\gamma^{tot}\rangle(t)$, with $^{235}$U being the steepest. Those differences can be attributed to differences in the fission fragment yields produced in the different reactions, as shown in Fig.~\ref{fig:YA}. The heavy fragment yields produced in the case of $^{235}$U and $^{239}$Pu thermal neutron-induced fission reactions are rather similar, with most noticeable differences appearing in the complementary light fragment region. Differences are largest with $^{252}$Cf (sf), including the region near mass 132, where prominent fragment isomers reside. 

The present study is limited to spontaneous and thermal neutron-induced fission reactions. At higher incident neutron energies, the symmetric region of the fission fragment yields, near 115-120 amu for fission of $^{236}$U* and $^{240}$Pu*, tends to fill up. Therefore isomers populated in this mass region would tend to appear more prominently in the late \gray~spectrum. However, as seen in Fig.~\ref{fig:lateGammasPerMass}, almost no late \grays~appear in this mass region in the case of $^{252}$Cf (sf), which does populate this mass region (see Fig.~\ref{fig:YA}). The situation is somewhat more complicated than what this intuitive argument would suggest, since at energies above multi-chance fission thresholds, more than one nucleus fissions with distributions of excitation energies that depend on the multi-chance fission probabilities and the compound plus pre-equilibrium components of the pre-fission neutrons being emitted.

\subsection{Dependence on Model Input Parameters}

\CGMF~calculations use several ingredients as either input data or model input parameters. The input data are the pre-neutron fission fragment yields in mass, charge, and kinetic energy, which are reconstructed from available partial experimental data. The fragment mass yields $Y(A)$ are model-dependent, as only the post-neutron emission fission fragment yields are measured, and assumptions on the prompt neutron emission have to be made to reconstruct the pre-neutron yields. In the case of the three well-known reactions studied here, this problem is certainly negligible. In addition, experimental data never come with a complete characterization of the fragments in mass, charge and kinetic energy simultaneously. Instead, one has to reconstruct the full correlated distribution $Y(A,Z,TKE)$ from partial information only. An uncertainty on the initial fission fragment yields would certainly impact some of the detailed results presented here. However, our predictions for the time evolution of average quantities such as the average total \gray~multiplicity and energy, as shown in Figs.~\ref{fig:Ngt-all} and~\ref{fig:Egt-all}, respectively, would remain robust.

Another input that could influence the results would be the use of a different nuclear structure database. For instance, if specific excited states and decay lines have not been observed experimentally, then \CGMF~calculations would not fill up the gap and would therefore not predict any isomeric state or decay. The ENSDF nuclear structure data library is updated regularly based on the most up-to-date experimental knowledge at a given time. On average again, the predictions shown in this paper should hold well, but specific decay schemes and particular \g~spectra for specific isotopes could certainly change over time, once an improved knowledge of their nuclear structures becomes available.

Beyond input data, \CGMF~also uses some model input parameters such as $R_T(A)$, which partitions the available total excitation energy between the two complementary fragments, and the global parameter $\alpha$ that scales up the moment of inertia of the fission fragments, thereby influencing their initial spin distribution. 

\begin{figure}[ht]
\centerline{\includegraphics[width=\columnwidth]{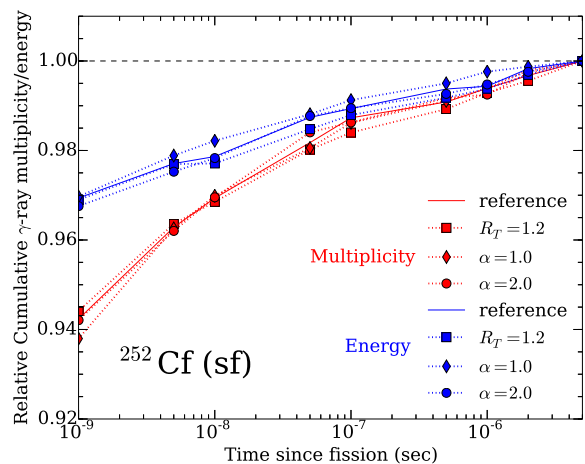}}
\caption{\label{fig:sensitivity}Sensitivity of the time dependence of the cumulative \gray~multiplicity (red) and energy (blue) on the choice of the model input parameters $R_T$ and $\alpha$. The ``reference" results, shown in Figs.~\ref{fig:Ngt-all} and~\ref{fig:Egt-all}, are reported here as solid lines.}
\end{figure}

In default \CGMF~calculations, the $R_T$ parameter depends on the mass of the fragments and is tuned to reproduce the saw-tooth curve representing \nubar$(A)$. The resulting curve for $R_T(A)$ can be fairly well reproduced by taking into account the deformation of the fragments near the scission point, and sharing the intrinsic part of the total excitation energy according to thermal equilibrium between the two fragments. Regardless of the validity of this very simple model, here we are interested in analyzing the sensitivity of our results concerning the time dependence of prompt \gray~emissions on the choice of $R_T(A)$. As an extreme case, we considered $R_T=1.2$, fixed for all mass splits. The result is shown in Fig.~\ref{fig:sensitivity} for both the cumulative \gray~multiplicity and energy, in the case of $^{252}$Cf spontaneous fission. Comparing this result to the ones shown for the same isotope and reaction in Figs.~\ref{fig:Ngt-all} and~\ref{fig:Egt-all} (also shown as solid lines on Fig.~\ref{fig:sensitivity}), one observes only small differences and the overall time dependences are very similar. Of course, results obtained with $R_T=1.2$ would not reproduce the saw-tooth curve and therefore does not represent a realistic situation. 

Similarly, changing the global $\alpha$ parameter increases or decreases the average initial angular momentum in the fission fragments. As a reference, \CGMF~calculations for $^{252}$Cf (sf) are done with $\alpha=1.7$. We have calculated the same results for $\alpha=1.0$ and 2.0. Results for 2.0 are expected to be closer to the reference ones, as indeed observed in Fig.~\ref{fig:sensitivity}. But even in the more radical change of $\alpha=1.0$, the trends are very similar. For $\alpha=1.0$, \CGMF\ would predict a harder prompt fission \gray~spectrum (see~\cite{Stetcu:2014}).

Note that these plots (Figs.~\ref{fig:Ngt-all}, \ref{fig:Egt-all} and \ref{fig:sensitivity}) show {\it relative} results only, normalized to unity at 5 $\mu$sec. The calculated absolute \gray~multiplicity and energy will differ for each case considered, although their relative time-dependent behaviors remain similar.

\section{Conclusion}

We have studied the time evolution of the emission of prompt \grays~in the thermal neutron-induced fission reactions of $^{235}$U and $^{239}$Pu, and in the spontaneous fission of $^{252}$Cf. Up to 3 to 7\% of all prompt \grays~are emitted between 10 nsec and 5 $\mu$sec following fission. These late prompt \grays~originate from the deexcitation of isomeric states populated in the post-neutron fission fragments. A proper description of these emissions is particularly important for studies of the prompt neutron spectrum and multiplicity. We have shown that modern Monte Carlo codes such as the one developed at LANL, \CGMF, can account for most of these time-dependent features reasonably well. \CGMF~makes use of the ENSDF nuclear structure database (through RIPL-3), which is continuously updated to account for new spectroscopic measurement, and in particular, of fission products. The number and energy of those late prompt \grays~depend on each isotope and fission reaction considered since they are strongly influenced by the pre-neutron emission fission fragment yields. While the present study concerns spontaneous and thermal neutron-induced fission reactions only, we have shown how changes in the population of particular fission fragments impact the time evolution of prompt fission \gray~data. At higher energies, the onset of multi-chance fissions and pre-equilibrium effects complicate this picture.

\section{Acknowledgments} 

We would like to acknowledge stimulating discussions with S. Oberstedt, A. Oberstedt, F.-J. Hambsch, A. G\"o\"ok, N. Carjan, M. Jandel, C. Walker, and A. Tonchev. This work was performed at Los Alamos National Laboratory, under the auspices of the National Nuclear Security Administration of the U.S. Department of Energy at Los Alamos National Laboratory under Contract No. DE-AC52-06NA25396. This work was partly supported by the Office of Defense Nuclear Nonproliferation Research \& Development (DNN R\&D), National Nuclear Security Administration, US Department of Energy.

\newpage
\newpage
\bibliographystyle{unsrt}

\end{document}